\documentstyle[preprint,floats,aps,epsf,epsfig,prb]{revtex}

\newcommand \ltdash{\raise-1.8pt\hbox{$\scriptscriptstyle |$}}

\newcommand \bea {\begin{eqnarray} }
\newcommand \eea {\end{eqnarray}}

\newcommand \dg {^{\dagger}}

\newlength{\bxwidth}\bxwidth=0.8\textwidth

\newcommand\prm[2]{$ $\vskip 2.4 truein
\centerline{\epsfig{file=#1,width=\bxwidth} }\vskip 0.5truein
\centerline{{\bf Fig.} #2}}
\newcommand\prk[2]{$ $\vskip 2.4 truein\vskip -2 truein
\centerline {\epsfig{file=#1,width=\bxwidth}} \vskip 0.5truein
\centerline{{\bf Fig.} #2}}
\begin{document}
\draft
\title{
Hidden Orbital Order in $URu_{2}Si_{2}$}
\author{P. Chandra,$^1$ P. Coleman,$^2$
J. A. Mydosh$^{3}$ and
V. Tripathi$^{2}$ }
\address{$^{1}$NEC, 4 Independence Way, Princeton, NJ 08540, U.S.A.}
\address{$^{2}$Center for Materials Theory, Dept of Physics and Astronomy, 
Rutgers University, Piscataway, NJ 08855, U.S.A. } 
\address{$^{3}$Kamerlingh Onnes Laboratory,  Leiden University,
P. O. Box 9504, 2300 RA Leiden, The Netherlands}
\maketitle
\vskip 0.35truein

{\bf 

When matter is cooled from high temperatures, collective instabilities
develop amongst its constituent particles that lead to new kinds of
order.\cite{Goodstein85}  An anomaly in the specific heat
is a classic signature of this phenomenon.  Usually the
associated order is easily identified, but sometimes its nature
remains elusive.  The heavy fermion metal $URu_2Si_2$ is one such
example, where the order responsible for the sharp specific heat
anomaly at $T_0=17 K$ has remained unidentified despite more than
seventeen years of effort.\cite{Buyers96}  In $URu_{2}Si_{2}$,
the coexistence
of large electron-electron repulsion and antiferromagnetic
fluctuations in $URu_2Si_2$ leads to an almost incompressible heavy
electron fluid, where anisotropically paired quasiparticle states are
energetically favored.\cite{Miyake86} 
In this paper we use these insights to develop a
detailed proposal for the hidden order in $URu_2Si_2$.  We show that
incommensurate orbital antiferromagnetism, associated with circulating
currents between the uranium ions, can account for the local fields
and entropy loss observed at the $17 K$ transition; furthermore we
make detailed predictions for neutron scattering measurements.  }

The intermetallic compound $URu_{2}Si_{2}$ contains a dense lattice of local
moments where quantum fluctuations barely prevent spin ordering;
the residual antiferromagnetic couplings between the 
strongly repulsive electrons are large and can drive new types of
collective instabilities.\cite{Buyers96} 
At $T_0 = 17.5 K$, $URu_2Si_2$ undergoes
a second-order phase transition characterized by sharp discontinuities
in bulk properties including specific heat,\cite{Palstra85} 
linear\cite{Palstra85} and nonlinear\cite{Miyako91,Ramirez92}
susceptibilities, thermal expansion\cite{deVisser86} and 
resistivity.\cite{Palstra86}  The
accompanying gap in the magnetic excitation spectrum\cite{Mason91} suggests
the formation of an itinerant spin density wave at this temperature;
however the size of the observed staggered moment\cite{Broholm91} ($m_0 = 0.03 \mu_B$)
{\sl cannot} account for the entropy loss at this transition.
The distinction between the primary hidden order parameter and
the secondary magnetic order parameter is clarified by high-field
measurements\cite{Mentink96,vanDijk97} which indicate that the bulk anomalies survive up
to an applied field strength of $40 T$, whereas the magnetically 
ordered moment is destroyed by fields less than half this magnitude ($15 T$).\cite{Mason95}  
Furthermore the size of the small ordered moment grows
linearly with pressure,\cite{Amitsuka99} while the gap associated with
the hidden order is relatively robust over this pressure range.\cite{Fisher90}  

There have been many theoretical proposals for the primary
order parameter in $URu_2Si_2$.\cite{Shah00}  
Two recent NMR studies have provided new insights on this
outstanding problem with crucial consequences
for the nature of the hidden order.  
It has been widely assumed
that the spin antiferromagnetism  and the hidden order are coupled
and homogeneous.
However a recent NMR study of $URu_2Si_2$
under pressure\cite{Matsuda01} indicates that for $T < T_0$
there exist {\sl distinct} antiferromagnetic and paramagnetic regions,
implying that the magnetic and the hidden orders are phase-separated.
This conclusion, supported by $\mu SR$ data,\cite{Luke94}
implies that the hidden order phase contains no spin order.
The observed growth of the staggered spin moment
with applied pressure\cite{Amitsuka99} is then simply a volume
fraction effect which develops separately from the hidden order
via a first order transition.\cite{Chandra01}  At ambient pressure
roughly a tenth or less of the system\cite{Matsuda01,Luke94} is magnetic with 
$m_{spin} = 0.3\mu_B$.  The key implication
of these measurements for theory is that the magnetic and the hidden order
parameters are independent.\cite{Chandra01}


In  a parallel study, NMR measurements
at ambient pressure\cite{Bernal01} on $URu_2Si_2$
indicate that at $T \le T_0$ the central (non-split) silicon NMR line-width develops a
field-independent, isotropic component whose temperature-dependent
magnitude is proportional to that of the hidden order parameter.
These results imply an isotropic field distribution at the silicon
sites whose root-mean square value is proportional to the hidden order
($\psi$)
\begin{equation}\label{facts}
\langle B^{\alpha } (i) B^{\beta} (j)\rangle = A^{2}\psi ^{2} \delta
_{\alpha \beta }, 
\end{equation}
and is $\sim 10$ Gauss at $T=0$.
This field magnitude is too
small to be explained by the observed moment\cite{Broholm91} which induces
a field
$B_{spin} = \frac{8\pi}{3}\frac{M}{a^3} = 100$ Gauss where
$a$ is the $U-U$ bond length ($a = 4 \times 10^{-8}$ cm).
Furthermore this moment is aligned along the $c$-axis,
and thus cannot account for the isotropic nature
of the local field
distribution detected by NMR.
These measurements indicate that as the hidden order develops,
an isotropically distributed static magnetic field develops
at each silicon site.  This is {\sl unambiguous} evidence
that the hidden order parameter breaks time-reversal invariance.

Guided by these recent experiments, we now discuss
our proposal for the nature of the hidden order parameter.
The magnetic fields at the silicon nuclei have two possible
origins:\cite{Schlichter78}  the conduction electron-spin interaction and the orbital
shift due to current densities.  In $URu_2Si_2$, the electron
fluid exhibits a strong Ising anisotropy along the $c$-axis,
as measured by the Knight shift;\cite{Bernal01} hence electron-spin coupling
cannot be responsible for the observed isotropic fields.
It would thus appear that these local fields are induced
by currents that develop inside the crystal as the hidden
order develops, and accordingly we attribute the observed isotropic
linewidth to the orbital shift.  

In this paper we propose
that $URu_2Si_2$ becomes an 
incommensurate orbital antiferromagnet (OAF) at $T = T_0$
with charge currents circulating between
the uranium ions.  Similar states have been studied extensively
in the context of the two-dimensional 
Hubbard model,\cite{Halperin68,Affleck88,Kotliar88,Nersesyan89,Schultz89}
particularly in connection with staggered flux phases.\cite{Hsu91,Wen96}
More recently commensurate current density wave order
has been proposed to explain the spin-gap phase in the
underdoped cuprate superconductors.\cite{Chakravarty01,Tewari01}
The planar tetragonal structure of $URu_2Si_2$ lends
itself naturally to an anisotropic charge 
instability of this type.
Here we show that incommensurate orbital antiferromagnetic
order in $URu_2Si_2$ can quantitatively account for the existing
specific heat and NMR
data for $T \le T_0$. 
In order to test this proposal, 
we make specific predictions for neutron scattering.
Calculation of the structure factor requires
knowledge of the fields throughout the full volume
in real-space.  Since the NMR only yields this information
at discrete points we need some additional input to proceed.
In the orbital antiferromagnet, the spatial dependence
of the fields throughout the sample is determined
by the requirement that the field distribution
at the silicon sites is isotropic.
Using this approach, we are able to link quantitatively 
the fields observed by NMR to the large specific heat anomaly
that develops at $T_0$.  We also use the spatial field
distribution associated with the incommensurate orbital
antiferromagnet
to predict the position, intensity and form-factor
associated with neutron scattering peaks on a surface of constant
anisotropy in momentum space centered on the origin.

We begin by estimating the local fields at the silicon sites
due to orbital currents circulating around the square uranium
plaquettes in the $a-b$ plane.  On dimensional grounds, the
current along the $U - U$ bond is given by 
$I \sim \frac{e\Delta}{\hbar}$ where $\Delta$ is the gap associated with 
the formation of hidden order. Then the field induced at
a height $a$ above a plaquette is 
$B \sim \left(\frac{2}{a \, c}\right)
\left(\frac{e\Delta}{\hbar}\right)$ = 11 Gauss, in
good agreement with the local field strength detected in $NMR$
measurements; here\cite{Palstra85,Mason91,Broholm91,Mentink96} we have used 
$\Delta = 110 K$.  The resulting 
orbital moment, $m_{orb} \sim  0.02 \mu_B$, is 
comparable to the effective spin moment
at ambient pressure ($m_0 = m_{spin} [P_{amb}] = [.10] m_{spin} = 0.03
\mu_B$).  
We emphasize that an orbital moment produces a local field
an order of magnitude less than that associated with a spin 
moment of the same value; the low field strengths observed at the
silicon sites are quantitatively consistent with our proposal
that they originate from charge currents.

An orbital moment, $m_{orb} = 0.02 \mu_B$, can also account for
the sizeable entropy loss at the transition.  In  a metal the change in
entropy is given by  $\Delta S= \Delta\gamma_{n}T_{0}$, where
$\Delta\gamma_{n}$ is  the change in the linear specific heat
coefficient resulting from the gapping of the Fermi surface. $\Delta
\gamma_n$  is inversely proportional to the Fermi energy
$\epsilon_{F}$ of the gapped Fermi surface.
Thus in general, the change in entropy per unit cell is given by
$\Delta S \sim k_{B}\frac{k_BT_{0}}{\epsilon _{F}}$. 
Exploiting the mean-field nature of this transition ($2\Delta \propto
T_0$), we find that 
\begin{equation}\label{orby}
\Delta{S}_{OAFM} \sim k_B \left(\frac{m_{orb}}{\mu^*_B}\right)=
0.02 k_B \left( \frac{\mu_{B}}{\mu^{*}_{B}}\right)
\end{equation}
where $\mu^*_B = \left(\frac{e}{\hbar c}\right) a^2 \epsilon_F$,
is the saturated orbital moment reached when the hidden order gap
$\Delta $ is approximately 
$\epsilon_F$. 
The ratio $\mu_{B}/\mu^{*}_{B}= (\frac{a_{o}}{a})^{2} (
\frac{\epsilon_{H}}{\epsilon_{F}})$, where 
$a_0$ is the Bohr radius and  $\epsilon_H= e^2/(2 a_0)$ 
is the energy of a hydrogen atom. For $URu_2Si_2$, the very large size
of $\left(\frac{\epsilon_H}{\epsilon_F}\right) \sim 10^3$ produced by
the large mass renormalization of the heavy electrons 
actually offsets the
small ratio $\frac{a_{0}}{a} \sim 10^{-1}$, so that
$\mu_{B}/\mu_{B}^{*}\sim 10$, and 
$\Delta
S_{OAFM} \sim 0.2 k_{B}=  0.3 k_{B} \ln 2$ is a number in 
good agreement with experiment.\cite{Palstra85}
The critical field for suppressing the associated
thermodynamic anomalies is
distinct from its spin counterpart; the
ratio 
$\frac{H_c^{orb}}{H_c^{spin}} \sim {\frac{\mu_B}{\mu_B^*}} \sim 10$
is qualitatively consistent with  the observed critical field
of approximately $40T$ associated
with the destruction of the hidden order\cite{Mentink96,vanDijk97,Mason95}.
From this simple discussion, we see that 
the sizable entropy loss associated with the development
of the  orbital antiferromagnetic state is a direct consequence of the
strong renormalization of the electron mass in $URu_2Si_2$ 
($\frac{m^*}{m} \propto \frac {\epsilon_H}{\epsilon_F}$).
The absence of such a large effective mass ($\frac{m^*}{m} \sim 3$)
in the cuprates could explain why analogous thermodynamic anomalies 
are difficult to observe there.  

Next we test whether the proposed orbital
antiferromagnetism will yield the observed
isotropic local fields.  In order to do so,
we allow the circulating
current around a plaquette (cf. Fig. 1)
centered at site $\bf  X$ to develop a modulated magnetization
$M({\bf  X})~ = ~\psi e^{i {\bf Q}\cdot {\bf  X}}$.
The current along a bond is then the difference of the circulating
currents along its adjacent plaquettes.
The field at a silicon site can be computed using Ampere's law, where
the relevant vector potential is
\begin{equation}\label{vpotential}
{\bf A} ({\bf x})= \frac{1}{c} \sum_j \int_{
{\bf x}_{j}^{(1)}
}^{
{\bf x}_{j}^{(2)}
}dx'\frac{{\bf I} ({\bf
x}_{j} )}{\vert {\bf x}- ( {\bf x}_{j}+{\bf  x}')\vert }.
\end{equation}
where ${\bf x}_{j}^{(1,2)}$ are the endpoints of the bond at site
${\bf x}_{j}$. 

The silicon atoms
in $URu_{2}Si_{2}$ are located at low-symmetry
sites, so that the fields do not cancel;
they reside above and
below the centers of the uranium plaquettes.
Microscopically, 
wave-vector selection is most likely due to details of the Fermi
surface, however we can obtain a good idea of the likely modulation
$Q$ vector from the 
isotropic nature of the field
distribution at the silicon sites. 
For example, in the commensurate case 
${\bf Q} = (1/2,1/2,1)$, the fields
on the silicon sites are only along the c-axis
and thus the field distribution would be highly anisotropic.
Consider a wave-vector $(q,q,1)$ (c.f. Fig. 1). In this case, the currents
are modulated within a plane with a wavelength $2 \pi /q$, but
staggered between  planes. This then produces a circulating
field distribution where the component of the field parallel
to the $(0,0,1)$ planes becomes larger and larger as $q$ is reduced.
To obtain an isotropic field distribution, $q$ needs to be reduced
to a point where the horizontal and vertical components of the field
are comparable.  A detailed calculation  based on the above model, shows that 
the condition of
perfect isotropy yields a circle of
wave-vectors (Fig. 2.) centered around $Q= (0,0,1)$ with a radius
$q\approx 0.22$. Relaxation of this constraint results in an annulus
of possible $Q$ vectors as shown in Fig. 2.

Our proposal of incommensurate current ordering
in $URu_2Si_2$ can be tested by experiment.
In particular we can Fourier transform the
real-space magnetic fields to calculate the
neutron scattering cross-section
\begin{equation}\label{S(q)}
\frac{d\sigma}{d\Omega} = \left( \frac{g_{N}e}{8 \pi \hbar c} 
\right)^{2}
|\vec{B}(q)|^2 =
r_o^2 S(q).
\end{equation}
Here $\vert \vec{B} (q)\vert^{2} $ is the structure factor of the magnetic
fields produced by the orbital currents and 
$S(q)= \vert B (q)\vert ^{2} / (4 \pi \mu_{B})^{2}$ is the  structure factor of the orbital magnetic moments,
measured
in units of electron Bohr magnetons ($\mu_{B}$). 
The factor 
$r_0 = \frac{g_N e^2}{4m_ec^2}$,  $m_e$ is the
electron mass and  $g_N$ is the neutron
gyromagnetic ratio. 
Using the vector potential from
eq. (3), we have calculated the magnetic field distribution
for the {\sl incommensurate } orbital antiferromagnet described above,
and find that its Fourier transform is given by
\begin{equation}\label{H}
{\vec{B}(q)}
=   { \frac{4\pi 
}{c}N I a^2} \hskip -0.15truein
 \sum_{{\bf G}\,n_1,n_2,n_3}
\hskip -0.15truein \delta_{q,Q+G}
        \ j_0\biggl[{\frac{q_x a}{2}}\biggr]j_0\biggl[{\frac{q_y
a}{2}}\biggr]      
         \ 
         (1 + (-1)^{(n_1 + n_2 + n_3)})
         \cdot \left(\frac{q_y{\hat x} - q_x{\hat
         y}}{2q}\right) \times {\hat q},
\end{equation}
where $j_0(x)=\frac{ \sin x}{x}$.
From this expression, we can determine the intensities
and the form factors associated with the diffraction.
We find that there exists a set of dominant peaks
associated with a
constant anisotropy locus of
wavevectors (Fig. 2.) in the first Brillouin zone (BZ)  where 
$S(q) = 0.18 \left(\frac{N Ia^2}{c \mu_B}\right)^2$,
indicating that roughly a fifth of the total integrated weight
of $S(q)$  (TIW) lies here.  Using the sum rule that relates
the TIW to the square of the moment, and the fact that at 
ambient pressure only a tenth of the sample is (spin) magnetic, we
find that the scattering peaks due to orbital ordering in the first
BZ should have 1/50 the intensity of the analogous spin magnetic peaks
at ambient pressure. 
We have also calculated that these
peaks will have a form-factor (cf. inset Fig. 2) that scales
with wavevector
as $q^{-4}$, where this power-law decay is
signatory of an extended scattering source. 

We end with remarks about the microscopic
formation of these charge currents. At low temperatures heavy
electron materials form almost incompressible Fermi liquids.  The large
Coulomb repulsion between the Landau quasiparticles
strongly suppresses on-site charge fluctuations, demanding
nodes in the particle-hole wavefunction.  The resulting anisotropically
paired states are also favored by the residual antiferromagnetic
interactions that persist in the heavy electron fluid.  Indeed
we believe that the same d-wave pairing that drives the
superconducting transition in $URu_2Si_2$ at $T=1.5 K$
also plays an important role in the formation of
the orbital antiferromagnetism.  We have found
that the development
of such anisotropic charge-density wave pairing occurs
naturally in a simple  model of $URu_2Si_2$ with
nearest-neighbor antiferromagnetic interactions, 
\begin{equation}\label{}
H_{I} = \sum_{\vec{q}}J (\vec{q})\vec{S} (\vec{q})\cdot \vec{S}
(-\vec{q}).
\end{equation}
Here $\vec{S} (\vec{q})$ is the Fourier transform of the
magnetization at wavevector $\vec{q}$ and 
$J (\vec{q})= 2 J (\cos q_{x}+ \cos q_{y})$ for a nearest-neigbour
interaction between adjacent $U$ atoms in the basal plane.
Expanding this interaction in terms of quasiparticle  
operators,
we find that it can be 
re-written as 
a sum of attractive interactions in four 
independent anisotropic charge density wave  channels:
\begin{equation}\label{33}
H_{I}= -J \sum_{\Gamma=1,4, k_{1},k_{2}, Q} (\gamma^{\Gamma}
_{k_{1}})^*\gamma^{\Gamma}_{k_{2}}\ \rho _{k_{1}} (Q)\rho _{k_{2}}
(-Q).
\end{equation}
Here $\rho_k(Q)= \sum_{\sigma =\pm 1/2 }c\dg_{k-Q/2\sigma }c_{k+Q/2\sigma }$ is the charge density
operator at wavevector $Q$, expressed in terms of creation and
annihilation operators  of the heavy quasiparticles.
The form factors are
$\gamma^{1,2} (\vec{k})= \cos (k_{x})\pm  \cos (k_{y}),
\gamma^{{3,4}}(\vec{k})= i(\sin (k_{x})\pm  \sin (k_{y}))$.
Of these four possibilities, channels $1$ and $3$ do not have nodes,
and are therefore suppressed by local Coulomb interactions. 
In the remaining two channels, only $\Gamma=4$ breaks
time reversal symmetry, giving rise to incommensurate orbital
currents.  One of the questions for further study concerns the
excitations of this mean-field state.  In particular, an
incommensurate OAF is expected to exhibit a gapless phason mode
associated with slow fluctuations of its wavevector.
$URu_{2}Si_{2}$ is known to develop a dispersing singlet 
excitation\cite{Broholm91}
at $T = T_0$. This mode was previously attributed to spin
antiferromagnetism, now known to be absent from the hidden order
phase.\cite{Matsuda01}  We plan to study  the possible
identification of this propagating mode with the phason of an incommensurate orbital
antiferromagnet.

In summary, we have discussed the theoretical implications of
two recent NMR experiments on the hidden order in $URu_2Si_2$.  Pressure-dependent
measurements indicate that it is completely independent
of the spin magnetism in this material.  We argue
that the development of isotropically distributed magnetic
fields at the silicon sites at $T=T_0$ implies that the
hidden order parameter breaks time-reversal symmetry.  
The size and the isotropy of these observed local
fields lead us to propose that $URu_2Si_2$ becomes an incommensurate
orbital antiferromagnet at $T< T_0$.
The  heavy electron mass reduces the saturation value of the orbital
moment, accounting for the sizable  entropy loss at the
transition and the scale of the associated critical field. We calculate the positions, intensities and
form-factor associated with the resulting neutron scattering peaks
so that this proposal can be tested by experiment.

We would like to acknowledge discussions with G. Aeppli, H. Amitsuka,
O. Bernal, S. Chakravarty, L.P. Gorkov, G. Lonzarich, K. McEuen, D. McLaughlin, D. Morr and C. Nayak.
P. Coleman and V. Tripathi are supported by the National Science
Foundation.

\vfill \eject


\newpage

\noindent{\bf Figure Captions}

\renewcommand{\labelenumi}{{\bf Fig.} \theenumi .}
\begin{enumerate}

\item 
Magnetic field distribution associated with incommensurate orbital 
currents in the $(0,0,1)$ plane.
(a) Schematic illustration of incommensurate orbital currents,
showing resulting magnetic fields above and below the $(0,0,1)$ plane,  
(b) Side view showing how proposed
field distribution is isotropic at silicon sites with a staggered
current
distribution between layers corresponding to $Q= (0.16,0.16,1)$.
\label{1}

\item 
An isotropic magnetic field distribution at the silicon sites can
be produced by many different wave vectors $Q$ for orbital order, all
of which give qualitatively similar neutron scattering patterns.    
(a) Contour plot showing locus of constant anisotropy 
around a ring of radius $Q_{\perp}=0.22$ in the vicinity of $Q= (0,0,1)$. 
(b) Predicted elastic neutron scattering intensity, where 
$\tilde{S} (q)^{\frac{1}{2}}= S
(q)^{\frac{1}{2}}/\left(\frac{NIa^{2}}{c\mu _{B}} \right)
\sim \sqrt{\vert B (q)\vert
^{2}}$
gives a measure of the Fourier spectrum of magnetic fields $B (q)$, 
plotted for the representative case $Q= (0.16,0.16, 1)$.

\label{2}

\label{3}

\end{enumerate}

\newpage


\prk{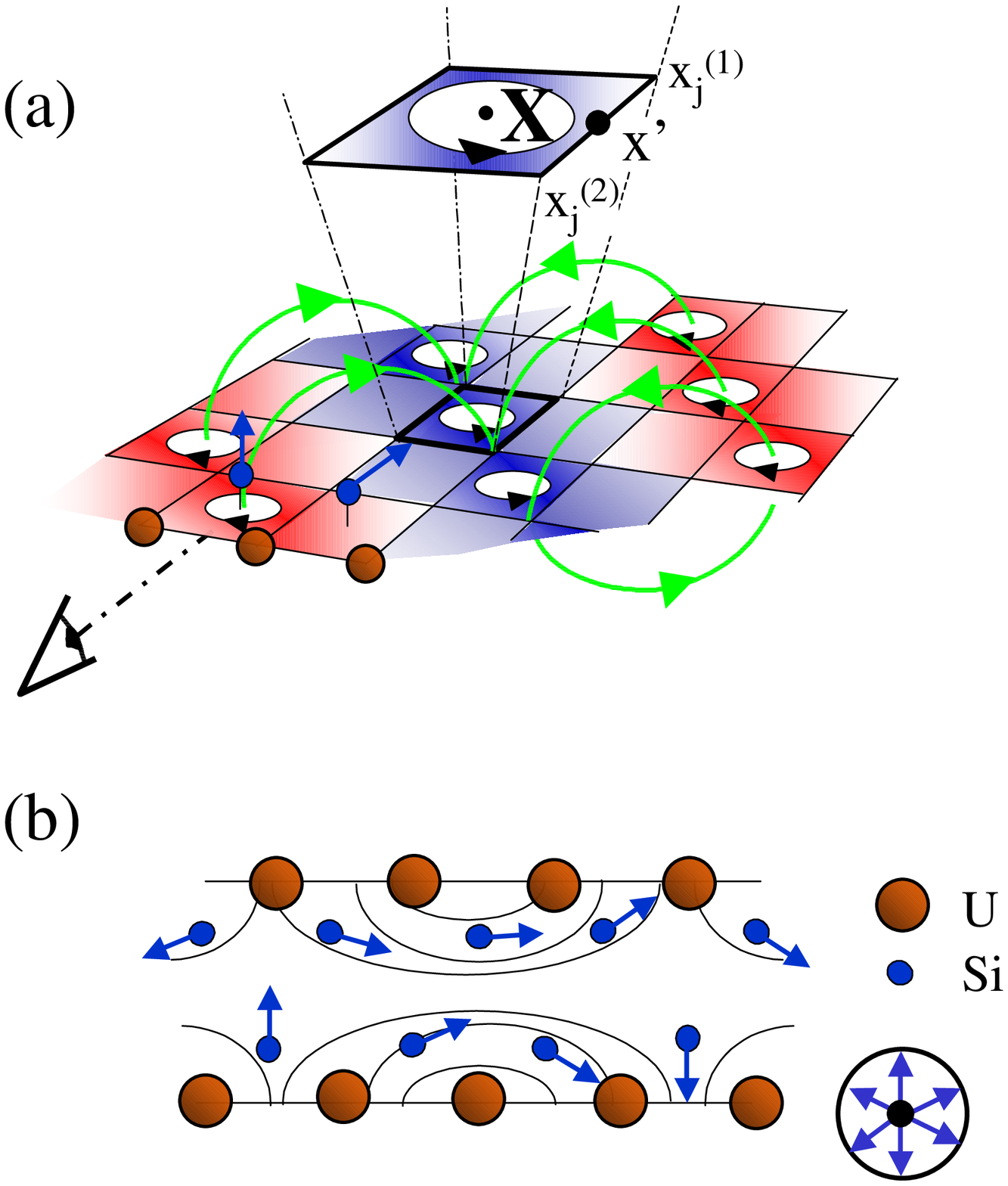}{1}



\prm{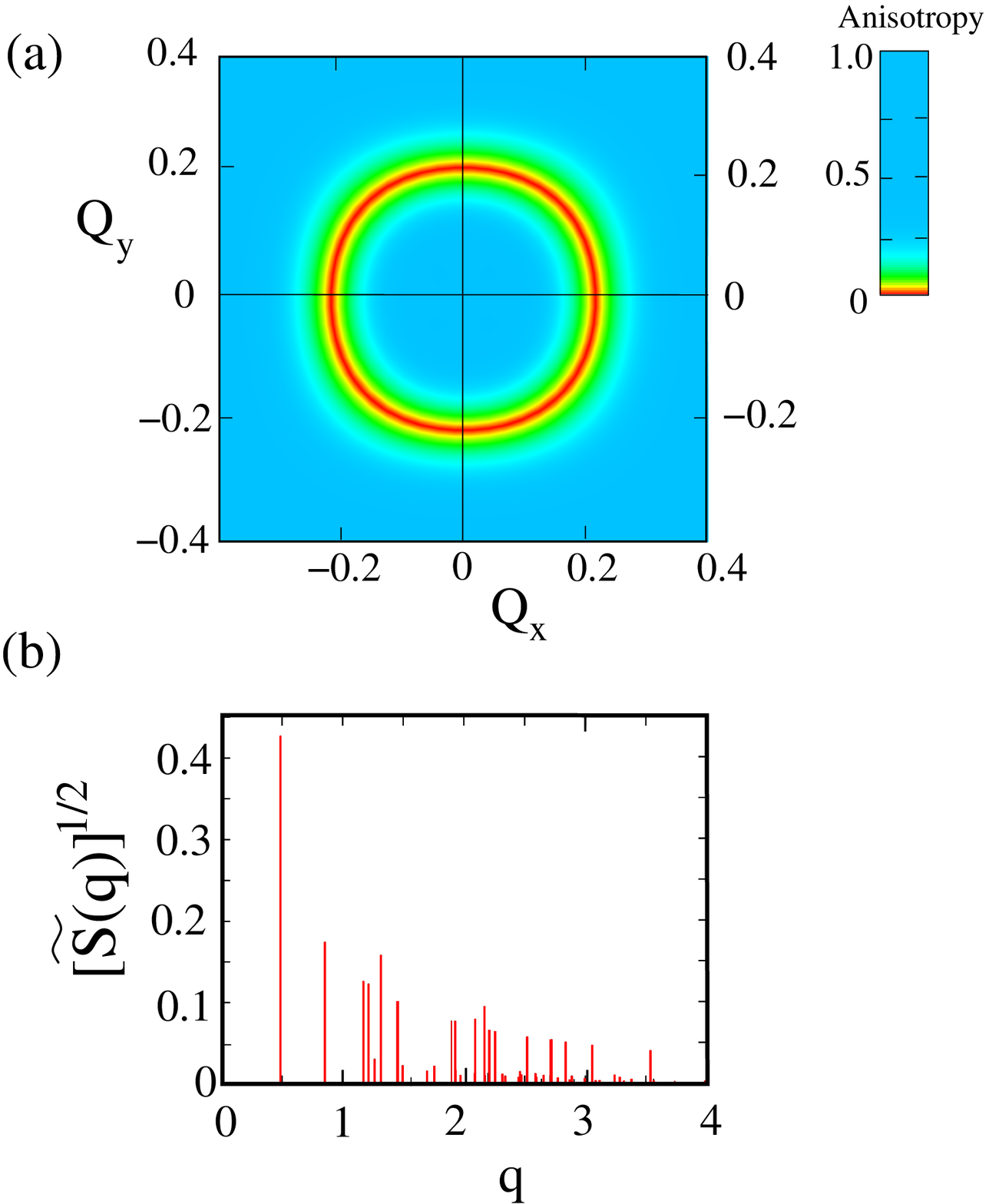}{2}




\end{document}